\documentclass[11pt]{article}

\usepackage{geometry}
\usepackage{currfile}
\usepackage{textcomp}

\usepackage{xcolor,soul}
\sethlcolor{yellow}

\usepackage[pdftex]{graphicx}
\DeclareGraphicsExtensions{.pdf,.jpeg,.png}
\graphicspath{{./.}}

\date{\today}

\begin{document}

\title{ \LARGE{Towards an Open and Scalable Music Metadata Layer}\\
~~\\
\large{ (DRAFT -- \today)  }  }
\author{
\normalsize{Thomas~Hardjono$^1$~~George~Howard{$^2$}~~Eric~Scace$^1$~~Mizan~Chowdury$^1$ }\\
\normalsize{{Lucas Novak}{$^1$}~~{Meghan Gaudet}{$^2$}~~{Justin Anderson}{$^1$}~~{Nicole~d'Avis}{$^2$} }\\
\normalsize{{Christopher Kulis}{$^2$}~~{Edward Sweeney}{$^2$}~~{Chandler Vaughan}{$^1$} }\\
\normalsize{~~}\\
~~\\
\normalsize{{$^1$}MIT Connection Science \& Engineering}\\
\normalsize{77 Massachusetts Avenue}\\
\normalsize{Cambridge, MA 02139, USA}\\
\normalsize{ {\tt hardjono@mit.edu};~{\tt scace@mit.edu};~{\tt mizanul@mit.edu} }\\
\normalsize{ {\tt ldnovak@mit.edu};~{\tt jander@mit.edu} }\\
~~\\
\normalsize{{$^2$}Berklee College of Music}\\
\normalsize{150 Massachusetts Avenue}\\
\normalsize{Boston, MA 02115, USA}\\
\normalsize{ {\tt ghoward@berklee.edu};~{\tt mgaudet@berklee.edu};~{\tt ndavis2@berklee.edu} }\\
\normalsize{~~}\\
}

\maketitle

\begin{abstract}
One of the significant issues in the music supply chain today is the lack of consistent, 
complete and authoritative information or metadata regarding
the creation of a given musical work.
In many cases multiple entities in the music supply chain have each created their own
version of the metadata for a musical work,
often by manually re-entering the same information or through scraping data from other sites.
In such cases, the effort to synchronize or to correct the information becomes manually laborious and error-prone.
Furthermore, confidential information regarding the legal ownership of the musical work
is often commingled in the same metadata,
making the entire database proprietary and thus closed.
In this paper we explore an alternative model for creation metadata
following the open access paradigm found in other industries,
such as in book publishing, library systems and in the automotive parts supply chain.
The vision is to create a new {\em music metadata layer}
for creation metadata that is open, scalable and provides
an authoritative source of information that is available
to all entities in the music supply chain globally.
\end{abstract}

\newpage
\clearpage

{\small 
%\par\noindent\rule{\textwidth}{0.4pt}
\tableofcontents
%\par\noindent\rule{\textwidth}{0.4pt}
}

%%%%%%%%%%%%%%%%%%%%%%%%%%%%%%%%%%%%%%%%%%%%%%%%%%%%%%%%%%%%%
\section{Introduction: Current Challenges in Music Metadata}

One of the significant issues in the music supply chain currently is the lack of consistent, 
complete and authoritative metadata regarding the creation of a given musical work.
In this paper we use the term {\em creation metadata} or simply ``metadata''
to denote the factual information regarding a given musical work (e.g. composition, sound recording) without
including the musical work itself (e.g. sound recording files).
Similar to other supply chains (e.g. containerized goods in shipping),
accurate information is needed about an item in order for entities across the supply chain
to be able to synchronize their business processes.
Today, in the data-driven society data is the new ``oil'' that drives digital ecosystems~\cite{WEF2011}.

Currently the music industry have created standards for metadata file formats (e.g. DDEX based on XML),
but the industry as a whole does not as yet have widely adopted standards which define
the processes or workflows by which creation metadata is collected, displayed and validated.
Often different parts or ``fractions'' of metadata are kept at different 
locations by different entities along the music supply chain~\cite{Deahl2019}.
Given that data accuracy is largely a solved problem
in other industries (e.g. financial industry) that employ technologies
based on advanced distributed databases
and tightly-synchronized transactional systems (e.g. NASDAQ, NYSE, etc.),
we believe that this music metadata problem should be the first and foremost challenge
the music industry needs to collectively address today.
This lack of standards for metadata workflows is only one of the many problems
plaguing the industry as a whole
(e.g. see~\cite{Howard2017a,Howard2017,Messitte2015}).

A note as to the notation used in this paper.
We use the general term of {\em musical work} to denote the individual song, composition or track, 
and we consider a song composition and recorded song as 
two separate musical works for the purposes of this paper. 
This holds true even for a song where the songwriter/composer 
and recording artist/performer are the same person. 
We use the term {\em creation metadata} to refer
to factual information regarding a given musical work.
The creation metadata must not include the actual musical work itself (e.g. sound recording MP3 or WAV)
and must not carry the legal ownership or copyright information
of the musical work.
This is akin to the bibliographic description of a book 
stored by the Library of Congress
and other libraries,
which does not include the book itself and 
does not include information on the current legal owner of the copyright of the book.
We use term {\em rights metadata} for information pertaining
to the legal ownership of rights to the musical work.
In many circumstances, rights-metadata may be considered
confidential, and thus in those cases it should not be made available to the public.
We use the term {\em distributed ledgers} (or simply ``ledger)
to denote the broader notion of blockchain systems and networks.
This allows the constructs in this paper to be implemented
using a variety of ledger implementations
(e.g. Ethereum~\cite{Buterin2014}, R3/Corda~\cite{R3-website}, Hyperledger~\cite{Androulaki2018SHORT}).

\begin{figure}[!t]
\centering
\includegraphics[width=0.9\textwidth, trim={0.0cm 0.0cm 0.0cm 0.0cm}, clip]{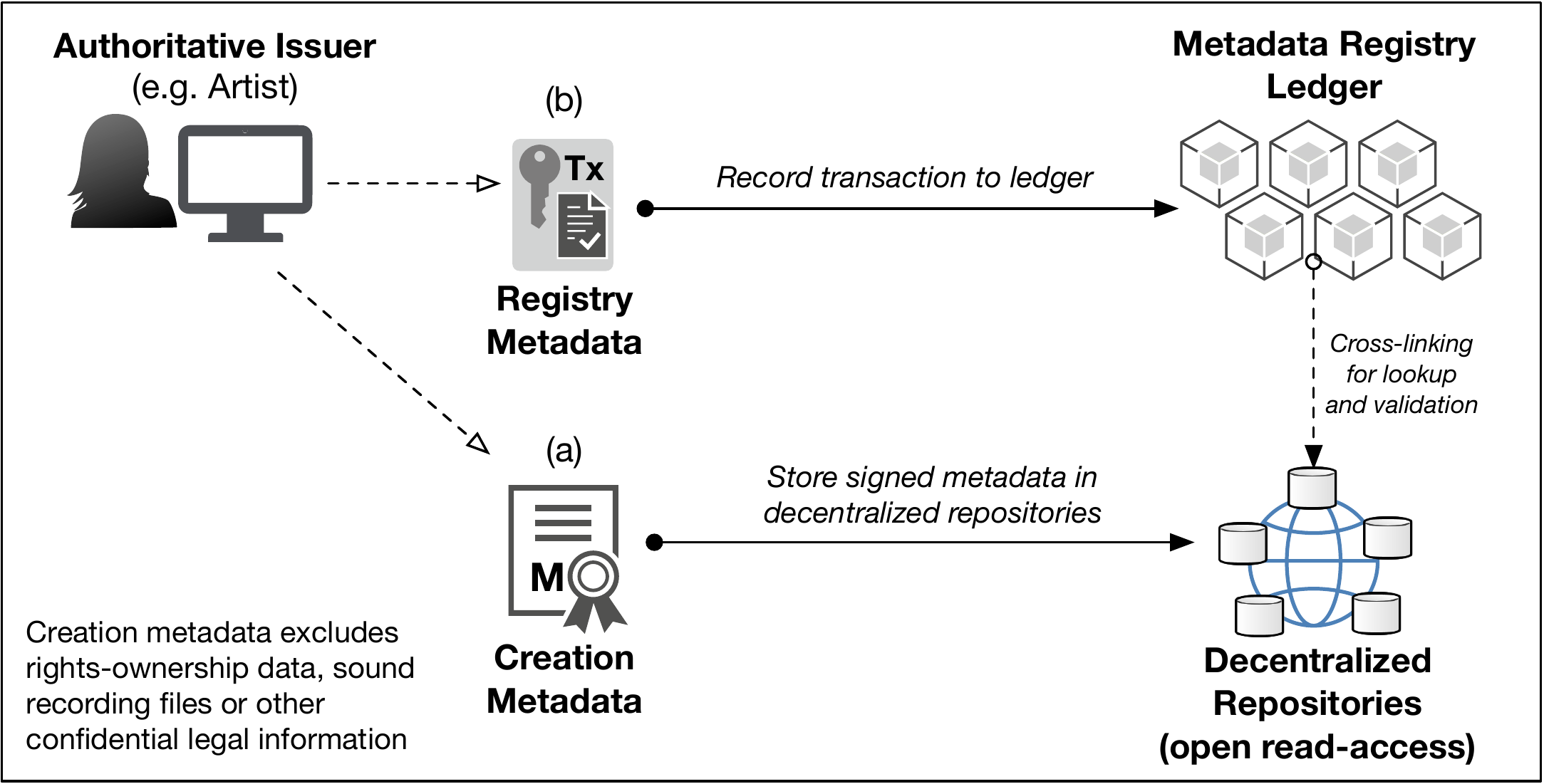}
	%
	% TRIMMING:  trim={<left> <lower> <right> <upper>} and clip options:
	% FULL EXAMPLE: \includegraphics[width=0.4\textwidth, trim={0.5cm 0.5cm 0.5cm 11.3cm}, clip]{image1.pdf}
	%
\caption{Summary of the core components of the Music Metadata Layer}
\label{fig:MetadataLayerComponents}
\end{figure}

The MIT Connection Science group and the Berklee College of Music
are leading the effort today to begin developing technical solutions
for standardizing the various constructs around an 
open access {\em music metadata layer}~\cite{HardjonoScace2019}.
The goal of the project is to explore the various technical issues
in creating an interconnected set of metadata repositories,
and to understand how this new open access music metadata layer can become the basis
for future music-related transactions on a distributed ledger or blockchain system.

Thus, for each atomic unit of metadata 
information for a given musical work (e.g. one song or track),
there should be exactly one authoritative creation-metadata file for that unit.
This authoritative creation-metadata file must be digitally signed,
and be publicly readable from multiple metadata repositories around the world.
Its digital signature provides a means to detect unauthorized modifications
to the file.
If a given musical work has several versions (e.g. original release, remix, etc),
then a separate creation-metadata file must be generated and signed for each.

The availability of a single authoritative creation-metadata file
for each musical work allows computing processes and systems
to operate based on unambiguous metadata.
When a computer program (e.g. traditional software; smart contract)
implements the licensing function of a musical work
to a licensee, the licensee (person or business entity)
can ``point'' the smart contract to the exact creation-metadata file of interest.
If a musical work has several versions (e.g. versions of sound recordings)
and if the licensee seeks to obtain a license for all of these versions,
then the licensee can point to each of the relevant creation-metadata file
corresponding to each of the versions.
As such, we believe this open access music metadata layer
is crucial for reducing the complexity of business transactions,
reducing the error rate due to mis-identification of musical works,
and thus reducing the overall cost of operations 
for all entities in the music supply chain.

In summary, we see this new music metadata layer as consisting of two (2) core components
(see Figure~\ref{fig:MetadataLayerComponents}):
\begin{itemize}

\item	{\em Replicated and decentralized open-access metadata repositories} (Figure~\ref{fig:MetadataLayerComponents}(a)):
The first component is a series of open access metadata repositories that store creation metadata,
without rights-ownership information and without 
copyrighted musical works (i.e. composition notes files or sound recording files).
Here ``replicated'' means that the same authoritative metadata file
must be available unmodified from multiple locations on the Internet.
It must be ``open access'' in the sense that anyone should be able 
{\em read} the metadata file (e.g. song metadata and album metadata),
but only songwriters or artists (or someone else they authorize)
are permitted to generate (write) new metadata files describing their musical works. 

A secondary part of the metadata repositories are the {\em distributed search databases}
that contain keywords, phrases and tags that are associated with given
creation metadata.
The goal of these search-databases is to allow anyone from anywhere in the world
to search for musical works based on freeform text (i.e. words, sentences),
and to create their own local cache of these keywords.

We discuss the need for these
replicated and decentralized metadata repositories in Section~\ref{sec:DistributedRepo}

\item	{\em Metadata registry distributed ledger} (Figure~\ref{fig:MetadataLayerComponents}(b)):
The second component is a metadata registry ledger
to which creators can register their creation metadata.
The entries of the ledger 
must include a globally unique resolvable identifier that
allows anyone to follow the linked identifier to a copy of the complete
creation metadata somewhere on the Internet.

We discuss the benefits of such a distributed ledger in Section~\ref{sec:RegistryBlockchain}.

\end{itemize}

%%%%%%%%%%%%%%%%%%%%%%%%%%%%%%%%%%%%%%%%%%%%%%%%%%%%%%%%%%%%%%%%%%%%%%%%%%%
\section{Design Principles}
\label{sec:DesignPrinciples}

\begin{figure}[!t]
\centering
\includegraphics[width=0.9\textwidth, trim={0.0cm 0.0cm 0.0cm 0.0cm}, clip]{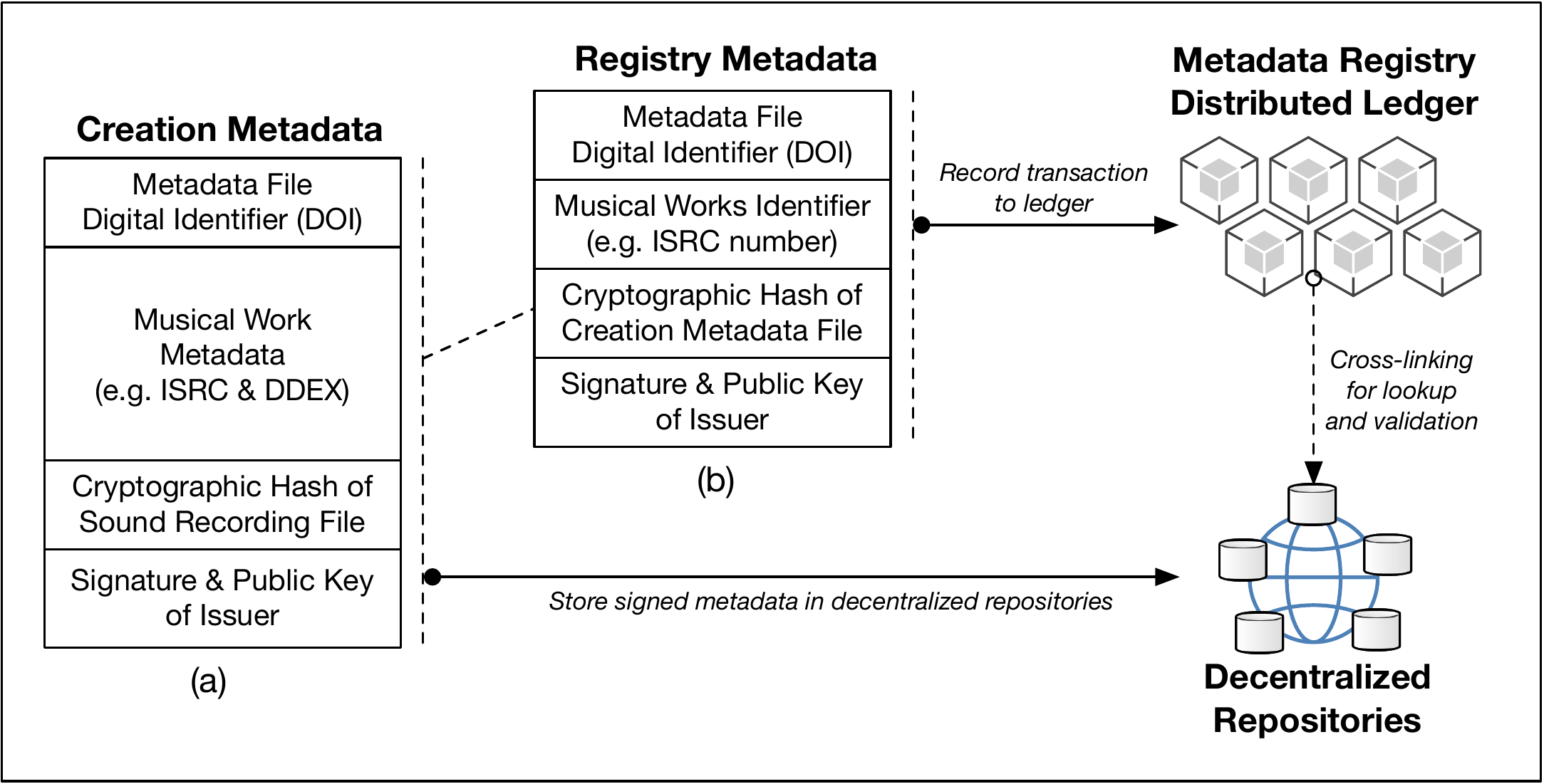}
	%
	% TRIMMING:  trim={<left> <lower> <right> <upper>} and clip options:
	% FULL EXAMPLE: \includegraphics[width=0.4\textwidth, trim={0.5cm 0.5cm 0.5cm 11.3cm}, clip]{image1.pdf}
	%
\caption{Summary of the creation metadata and registry metadata}
\label{fig:blockchainregistration}
\end{figure}

There are several design principles that should be the foundation for 
the technical architecture of the music metadata layer:
\begin{itemize}

\item	{\em Data collection upstream at point of works creation}:
Artists, musicians and relevant production-side entities
(e.g. studio engineers, producers, managers, etc.)
need to be empowered with the correct and intuitive tools (e.g. softwares)
to capture the creation information into a metadata file
and to digitally sign it (locally) 
as a means to assert a claim of ``authority'' over the provenance of that metadata.
Existing systems, such as Digital Audio Workstations (DAW),
may be a suitable point in the supply chain at which
factual information regarding the creation event can be captured.

\item	{\em Separation of creation metadata from rights-bearing musical works}:
Musical works (e.g. compositions, sound recordings)
must be separated from creation-metadata files for privacy reasons
and copyright enforcement reasons.

Several access control models, mechanisms and solutions exist in the market
today to provide protected access to these valuable resources
(e.g. OAuth2.0~\cite{rfc6749}, OpenID-Connect~\cite{OIDC1.0},
UMA~\cite{UMACORE1.0,UMACORE2.0}).

\item	{\em Separation of ownership information from factual creation metadata}:
Ownership information should be separated from creation metadata.
This is because the ownership information may be confidential and
because often the ownership to a given musical work may be sold/acquired over time.
Any change of ownership must not alter the creation metadata.

\item	{\em Open data access philosophy to creation metadata}:
The music metadata layer should adopt an {\em open access philosophy}
which is already common today in other industries and sectors.
Many publications today (e.g. books, magazines, journals, etc.)
have adopted the open access philosophy
for the purpose of advancing knowledge for the entire human race~\cite{openaccess-Wiki,Suber2012}.

Private contracts and confidential information should not be placed in
these public open access metadata repositories.
Similarly, the actual musical works (e.g. sound recording master files)
should not be placed in open access locations.

This open access philosophy for music metadata
paves the way towards a more accurate attribution of musical works to artists, 
musicians and other relevant creation-side persons and entities.
Additionally, an open access music metadata layer allows fans to obtain more details
about the creation of the musical work
(e.g. what type of keyboard the musician was using).
The availability of keywords and phrases linked to metadata files
allows for {\em intelligent search} capabilities to be developed atop the music metadata layer.

\item	{\em Inclusion of cryptographic hash of musical work}:
Every creation metadata file must include a cryptographic hash of the digital representation
of the musical work of concern
(e.g. hash of MP3 sound recording file) as a reference to the work.
This allows for an exact 1-to-1 mapping between the metadata file
and the corresponding musical work.

\item	{\em Standard descriptor for metadata format \& encoding}:
We anticipate that in the future there will be multiple
metadata formats (e.g. DDEX-XML~\cite{DDEX-website}, JSON) 
with various encodings (e.g. Unicode, Chinese~GB, etc),
according to the community of creators.
The relevant headers of every metadata file must include
such information in order to aid the reader (i.e. Client software)
that fetches and parses these metadata files.

\item	{\em Digitally signed and portable metadata units}:
The atomic unit of metadata (e.g. song, track, or composition as the atomic unit) 
must be digitally signed
by an authoritative entity,
such as the artist, songwriter, composer, musician, producer or other persons
who have been authorized to do so.
This allows the metadata unit
to be self-contained and stand-alone, and therefore be portable and replicable as a unit.
Each metadata file must carry a globally unique digital identifier that allows
it to be uniquely distinguished from other metadata units.

Figure~\ref{fig:blockchainregistration}~(a) illustrates this notion
of a signed metadata file with a unique file identifier,
where the last part in Figure~\ref{fig:blockchainregistration}~(a)
shows the signature portion 
(e.g. {X.509}~\cite{RFC2560-formatted,ITU-X509} or XML-DSig~\cite{W3C-Dsig-2015})
and the signer information.

\item	{\em Replicated copies of metadata units}:
The signed metadata files must be made available
to the public at multiple locations throughout the Internet.
The digital identifier used for the metadata file must allow
a user to locate one of the many copies of the metadata files
on the Internet.
Access should be made via standardized APIs.

\item	{\em Globally unique and resolvable file identifiers}:
Each metadata file must be assigned a unique identifier under a registered namespace
that allows a user to fetch a copy from one of the many repositories around the world
based on that identifier.

Digital identifier schemes such as the Digital Object Identifier (DOI)~\cite{DOI-ISO-Standard} and
its accompanying Handle resolver system~\cite{RFC3650,RFC3651} 
have been successfully deployed at a wide scale for over a decade now.
Similar in protocol-behavior to the DNS infrastructure,
the DOI and Handle allows for efficient look-ups
of copies of data files (e.g. library catalog entries) stored at open repositories all over the Internet.

\item	{\em Ledger-based notarization of signed metadata}:
A distributed ledger can be used as
a means to {\em notarize} each signed metadata unit or a shorter summary of it.
For convenience, we refer to the shorter metadata as the {\em registry metadata},
implying that the shorter registry metadata is intended to be stored on the distributed ledger.
The same file identifier (e.g. DOI) used in the creation-metadata
must be used in the registry-metadata,
signifying that both point to the same musical work.
The notarization of the registry metadata
provides a tamper-detectable timestamp on the musical work.

Figure~\ref{fig:blockchainregistration}~(b) illustrates the notion
of notarization via a {\em metadata registry ledger},
and shows the shorter registry metadata
to be recorded on the ledger.
Entities who fetch a version of a metadata file must check
the ledger to ensure that they have the latest version
(e.g. based on the timestamp on the confirmed transaction).

\item	{\em Support for multiple versions of a musical work}:
In many circumstances artists and musicians may create different versions of a given
musical work.
For example, for a given sound recording
an artist may record a short version (e.g. 2 minutes long), 
long version (e.g. 3 minutes long)
and an extended version (e.g. 6 minutes long).
Since the cryptographic hash of the musical work(e.g. MP3 file) is included
in the metadata file (see Figure~\ref{fig:blockchainregistration}~(a)),
this implies that a separate metadata file must be created for each of these versions.
Furthermore, this means that each of these metadata files must have a unique identifier.
This is crucial in order to allow a licensee to unambiguously point to the exact
version of the sound recording for which they are seeking a license.

\item	{\em Support for revisions of metadata}:
It is inevitable that errors may exist in metadata information
even when the information is collected upstream at the creation-end of the supply chain (e.g. in DAW software).
As such, if a metadata file is to be corrected or revised,
then the existing (old) metadata file should never be deleted or erased.
When a new revision is written,
the new version metadata file must be assigned a new file-identifier
and must contain a link to (e.g. hash of) the revised version.
This indicates to the reader (i.e. Client software) that a previous version exists.
This principle is akin to source {\em version-control} in software engineering
development, which is very common today (e.g. SVN or GitHub).

Similarly, when a revised version is to be notarized via the registry ledger,
the new registry-metadata must include a pointer to (e.g. hash of)
to the previously confirmed registry metadata
(e.g. transaction-id of previously confirmed registry-metadata transaction).

\item	{\em Support for archival of revised metadata}:
Revised (old) metadata files must never be erased,
and must be archived using the same replicated repositories architecture we have discussed. 
Archived metadata files must remain open access
in order to support the tracing the provenance and history of the metadata information.

\end{itemize}

There are at least three benefits in this approach of combining
open access metadata files and the notarization using the metadata registry ledger:
\begin{itemize}

\item	{\em Support of copyright claim}: 
The ``publishing'' of the signed registry-metadata file of a musical work
onto the registry ledger
provides legal support to the copyright claim on the part of the creator(s).
Furthermore, the notarization using a distributed ledger
(with an additional act of signing the transaction)
provides a relatively immutable and non-repudiable
timestamped public evidence of the existence of the musical work.

\item	{\em Basis for music metadata oracle for other ledgers \& systems}:
By being the open registry for musical works metadata,
the registry ledger effectively becomes the {\em oracle} for metadata
that can be referred to (linked to) by other types of ledger-based transactions,
such as license request smart contracts, and license granting smart contracts.
Even existing systems (e.g. legacy databases, or future systems such as that of 
the Mechanical Licensing Collective or the Repertoire Data Exchange)
can thus refer to the relevant entries (e.g. transaction-id) in the registry ledger.

\item	{\em Opportunity for tight binding between a digital identity and public key}:
The need to digitally sign the creation-metadata and to sign the transaction
submitted to the registry ledger necessitates dealing with key management
of the private-public key pair of the signer.
Standards for the creation of digital certificates
based on the strong identification of persons have already existed in
industry for over two decades now~\cite{HousleyPolk2001}.
New efforts to retain this binding in a decentralized fashion via
a blockchain system is also underway~\cite{W3C-DID-2018}.

This in turn paves the way towards solving the current problem
in the music industry of identifying rights-holders (persons or legal entities)
to whom royalty payments are owed (e.g. see~\cite{Messitte2015}).

\end{itemize}

%%%%%%%%%%%%%%%%%%%%%%%%%%%%%%%%%%%%%%%%%%%%%%%%%%%%%%%%%%%%%%%%%%%%%%%%%%%
\section{Distributed Repositories}
\label{sec:DistributedRepo}

\begin{figure}[!t]
\centering
\includegraphics[width=0.8\textwidth, trim={0.0cm 0.0cm 0.0cm 0.0cm}, clip]{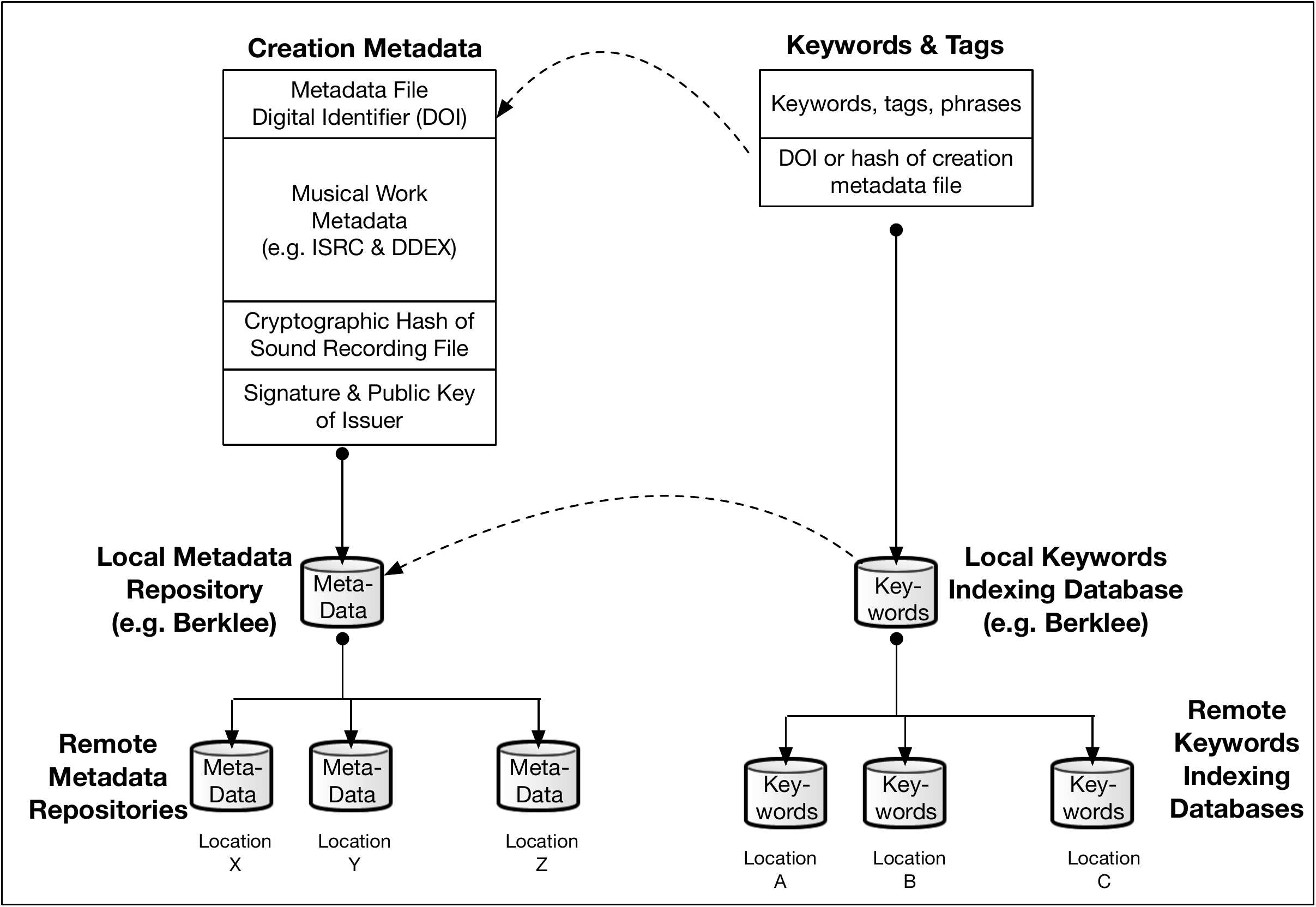}
	%
	% TRIMMING:  trim={<left> <lower> <right> <upper>} and clip options:
	% FULL EXAMPLE: \includegraphics[width=0.4\textwidth, trim={0.5cm 0.5cm 0.5cm 11.3cm}, clip]{image1.pdf}
	%
\caption{Replicated metadata repositories with linked keywords/tags indexing}
\label{fig:Replicatedrepositories}
\end{figure}

We envisage that the creation-metadata of a given musical work
should be replicated for higher-availability and reliability.
There are several basic requirements for the implementations
of the replicated metadata repositories:
\begin{itemize}

\item	{\em Independence of replication technologies}:
Since database and replication technologies will continue to evolve over time,
the creation-metadata as the atomic unit must be ``movable'' (copyable)
from one repository implementation to another.

\item	{\em Standardized Service APIs for metadata repositories}: 
The repository RESTful APIs used by a Client application that reads/writes to
a metadata repository must be standardized.
The API definitions must be independent
of the technological implementation of the repository behind the APIs.
A Client application calling to the APIs of a service should not need to be aware of
the backend implementation of the service.

\item	{\em Standardized Import \& Export data format retaining signatures}: 
When a metadata file is to be exported (read) from a metadata repository,
then the file must use a standardized data-format.
The digital signature portion of the original metadata file (as submitted initially by its creator)
must remain valid when the metadata file is exported.

For example, consider the case where an artist employs
a Client application (e.g. DAW) to write a new a creation-metadata file
in a given format (e.g. JSON) that is then signed 
by the artist
using the corresponding public-key signature standard (e.g. JSON Web Signature).
When the artist via the Client application submits this file to the local
metadata repository,
the repository may internally dissemble the metadata file
in accordance to its internal data storage architecture.
However,
when later a home-user reads (exports) a copy of the metadata
from this open access repository,
the repository must be able to re-assemble the original creation-metadata
such that its original signature can be validated by the home-user.

\end{itemize}

\begin{figure}[!t]
\centering
\includegraphics[width=0.9\textwidth, trim={0.0cm 0.0cm 0.0cm 0.0cm}, clip]{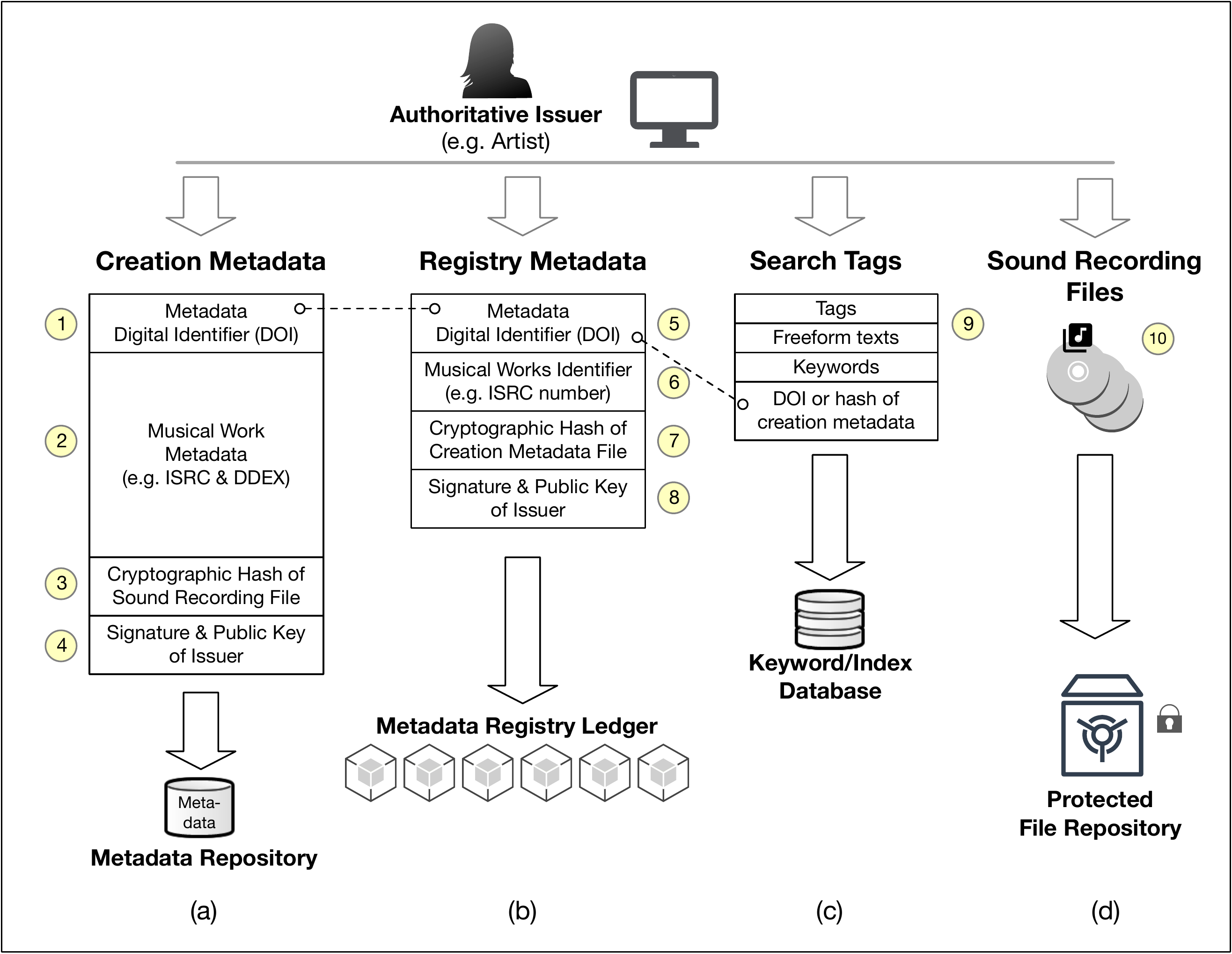}
	%
	% TRIMMING:  trim={<left> <lower> <right> <upper>} and clip options:
	% FULL EXAMPLE: \includegraphics[width=0.4\textwidth, trim={0.5cm 0.5cm 0.5cm 11.3cm}, clip]{image1.pdf}
	%
\caption{Summary of parts of (a) the creation-metadata and (b) the registry-metadata}
\label{fig:MetadataRegistration}
\end{figure}

Figure~\ref{fig:MetadataRegistration}(a) illustrates 
a summary of the parts of the creation metadata file:
\begin{itemize}

\item	Part~1: {\em Metadata Digital Identifier}: The first part is the
identifier of the creation-metadata file.
We propose to use the Digital Object Identifier (DOI) scheme~\cite{DOI-ISO-Standard,RFC3650,RFC3651}
due to its long history of successful deployments around the world.

\item	Part~2: {\em Musical Works Metadata}: The second part
pertains to the actual musical works metadata.
This component may use existing music metadata formats (e.g. XML-based DDEX RIN, JSON)
or it may use other formats.
The header part of this component must indicate the
format and encoding of the metadata.
Note that the metadata file must not contain the musical sound recording file
or legal ownership information.

\item	Part~3: {\em Cryptographic hash of musical works file}:
The third part in the creation-metadata file is a cryptographic hash 
of the musical work file.
For example, this could be a hash of a sound recording master file (e.g. MP3, MPEG4)
or the hash of the music notation file
(e.g. PDF file, Sibelius file, Finale file).

This allows for a correct 1-to-1 mapping between the metadata file and the musical works file.
This exact matching becomes relevant in business processing
when there are multiple versions
of a given musical work (e.g. long version, short version, or remix of a sound recording).

\item	Part~4: {\em Authoritative issuer digital signature}:
The issuer of the creation-metadata must digitally sign the combined parts
of the metadata (i.e. Part~1 to Part~3 above).

The digital signature is performed
using the existing standard techniques for public-key digital signatures and timestamps.
The signature portion is the fourth part 
of Figure~\ref{fig:MetadataRegistration}(a).
The signature part (Part~4) must include
the standard pieces of information
required for a reader to validate the file
(e.g. signature algorithm-ID, etc. -- see~\cite{HousleyPolk2001}).

Henceforth,
any attempt to modify any data in the assembled parts 
(Part~1 to Part~4 of Figure~\ref{fig:MetadataRegistration}(a))
will cause the signature-verification to fail -- indicating
to the reader that the creation-metadata file is no longer authentic
(i.e. that it has been tampered with).

\end{itemize}

%%%%%%%%%%%%%%%%%%%%%%%%%%%%%%%%%%%%%%%%%%%%%%%%%%%%%%%%%%%%%%%%%%%%%%%%%%%
\section{Metadata Registry Ledger}
\label{sec:RegistryBlockchain}

The music metadata layer may benefit from the use
of one or more distributed ledger networks
as a means to effect a simple consensus-based notarized {\em registry ledger}.
Since a creation-metadata file might be large
and given that most ledger-based transactional systems
are not intended to store large files,
only a short summary of the metadata
should be recorded on the registry-ledger system.
We refer to this as the {\em registry-metadata} structure
(see Figure~\ref{fig:MetadataRegistration}(b)).

The shorter registry-metadata will be recorded
on the registry ledger,
and its resulting transaction-ID on that ledger
can later be used as a reference in business logic implementations
and multi-party transactions
in other systems and ledgers.
The registry-metadata must always carry the same identifier (i.e. DOI) as the creation-metadata,
indicating that these two data structures refer to the same musical work.

\begin{figure}[!t]
\centering
\includegraphics[width=0.8\textwidth, trim={0.0cm 0.0cm 0.0cm 0.0cm}, clip]{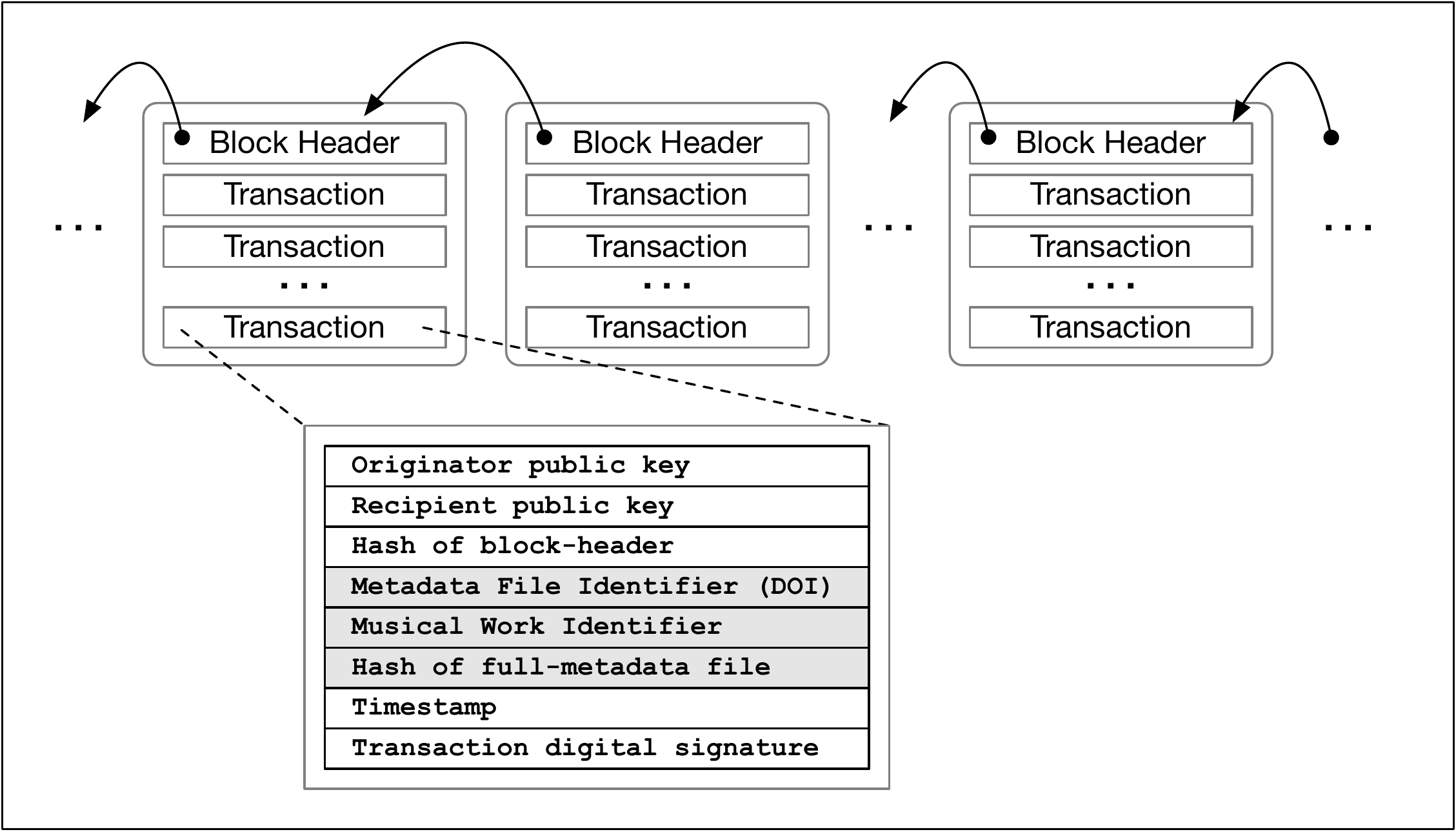}
	%
	% TRIMMING:  trim={<left> <lower> <right> <upper>} and clip options:
	% FULL EXAMPLE: \includegraphics[width=0.4\textwidth, trim={0.5cm 0.5cm 0.5cm 11.3cm}, clip]{image1.pdf}
	%
\caption{Illustration of a transaction in a block containing registry-metadata (grey fields)}
\label{fig:bc-transactions}
\end{figure}

There are two major goals of the registry ledger:
\begin{itemize}

\item	{\em Multicopy registry-metadata file and look-up}: The registry ledger
provides multiple copies of the registry-metadata, by virtue of the P2P network of nodes.
Each node independently keeps 
a full set of confirmed blocks of transactions (see Figure~\ref{fig:bc-transactions}),
each of which carries the registry-metadata structure.
The metadata identifier part (Part~5 in Figure~\ref{fig:MetadataRegistration}(b))
includes the same file identifier (e.g. DOI) as in the creation-metadata
(Part~1 in Figure~\ref{fig:MetadataRegistration}(a)).
This allows any entity to use the DOI identifier
found in the short registry-metadata (on the public registry ledger)
to fetch a copy of the full creation-metadata file from one or more
repositories on the Internet.

\item	{\em Persistent on-chain record for other infrastructures}:
The metadata registry ledger provides a persistent evidence
upon which
other infrastructures and systems can rely on (i.e. can point to).
Thus, a music licensing scheme implemented on a different ledger or blockchain system
can ``point to'' a registry-metadata found on the registry ledger
as part of a license issuance smart contract.
Similarly,
legacy systems and databases can cite the transaction-ID (on the ledger)
of the registry-metadata in its business logic processing software.

\end{itemize}

The parts of the registry-metadata is shown in Figure~\ref{fig:MetadataRegistration}(b)
(which does not show the enveloping ledger transaction structure):
\begin{itemize}

\item	Part~5: {\em Metadata Digital Identifier}: The first part is the 
digital identifier of the registry-metadata file, 
which must be identical to the value found in the full creation-metadata 
(first part of Figure~\ref{fig:MetadataRegistration}(a)).

\item	Part~6: {\em Musical Works Identifier}: The second part 
carries the musical works identifier 
that maybe in use within the metadata file. 
Typically this would be an identifier common and 
understood in the music industry (e.g. ISRC number for recordings or ISWC number for compositions).

\item	Part~7: {\em Cryptographic hash of full-metadata file}: The 
third part carries the cryptographic hash of 
the full-metadata file as a means to ensure 1-to-1 correspondence 
between the registry-metadata and the full creation-metadata file.

\item	Part~8: {\em Authoritative issuer digital signature}: The fourth part carries the 
digital signature of the authoritative issuer, 
which should be the same issuer as that of the full creation-metadata file.
Although not shown,
typically a timestamp is included in the digital signature data structure.

\end{itemize}
After the authoritative issuer (e.g. artist) completes
the capturing of the musical works metadata,
he or she proceeds to create a short registry-metadata
which is then enveloped within a transaction structure and transmitted to the distributed ledger.
The transaction's recipient (address) is either the issuer itself (i.e. to the public-key of the issuer)
or is ``null'' (depending on the specific ledger implementation in question).
This self-addressed transaction implicitly indicates
that it is a notarization transaction.

%%%%%%%%%%%%%%%%%%%%%%%%%%%%%%%%%%%%%%%%%%%%%%%%%%%%%%%%%%%%%%%%%%%%%%%%%%%
\section{Distributed Search and Lookups}
\label{sec:Distributed Search}

Another issue closely tied to the music metadata layer relates
to the ability for a user (e.g. home user, other creative artists) 
to search the various metadata repositories for music
using keywords and phrases.
We believe a separate ``search infrastructure'' is needed that is
parallel and interconnected to the various metadata repositories.

There are a number of interesting considerations for this search infrastructure 
based on an interconnected set of the keywords-databases shown earlier in 
Figure~\ref{fig:Replicatedrepositories} and Figure~\ref{fig:MetadataRegistration}(b).
Some of these considerations are as follows:

\begin{itemize}

\item	{\em Separation of creation metadata from search-material}:
The words, tags and phrases information -- referred to here as {\em search material} --
must be stored and managed separately from creation-metadata files.
This is because while the creation metadata may 
remain static over time (unchanged) once it has been signed,
the accumulated size of search-material -- consisting of 
permutations and combinations of words and phrases --
may grow and change over time.

\item	{\em Creator-side association of keywords and phrases}:
Artists and musicians must be able to associate their own words, tags and phrases
to a given music metadata and have this search-material
be stored locally, but be read-accessible globally.

\item	{\em User-side association of keywords and phrases}:
Similarly, any user or person (or AI and Machine Learning systems) must be able
to create their own association of words, tags and phrases
for a given music metadata and have this search-material
stored locally to them.
This is analogous to the current practice of music playlists
that users create on their devices and streaming-accounts.

\end{itemize}

\begin{figure}[!t]
\centering
\includegraphics[width=0.9\textwidth, trim={0.0cm 0.0cm 0.0cm 0.0cm}, clip]{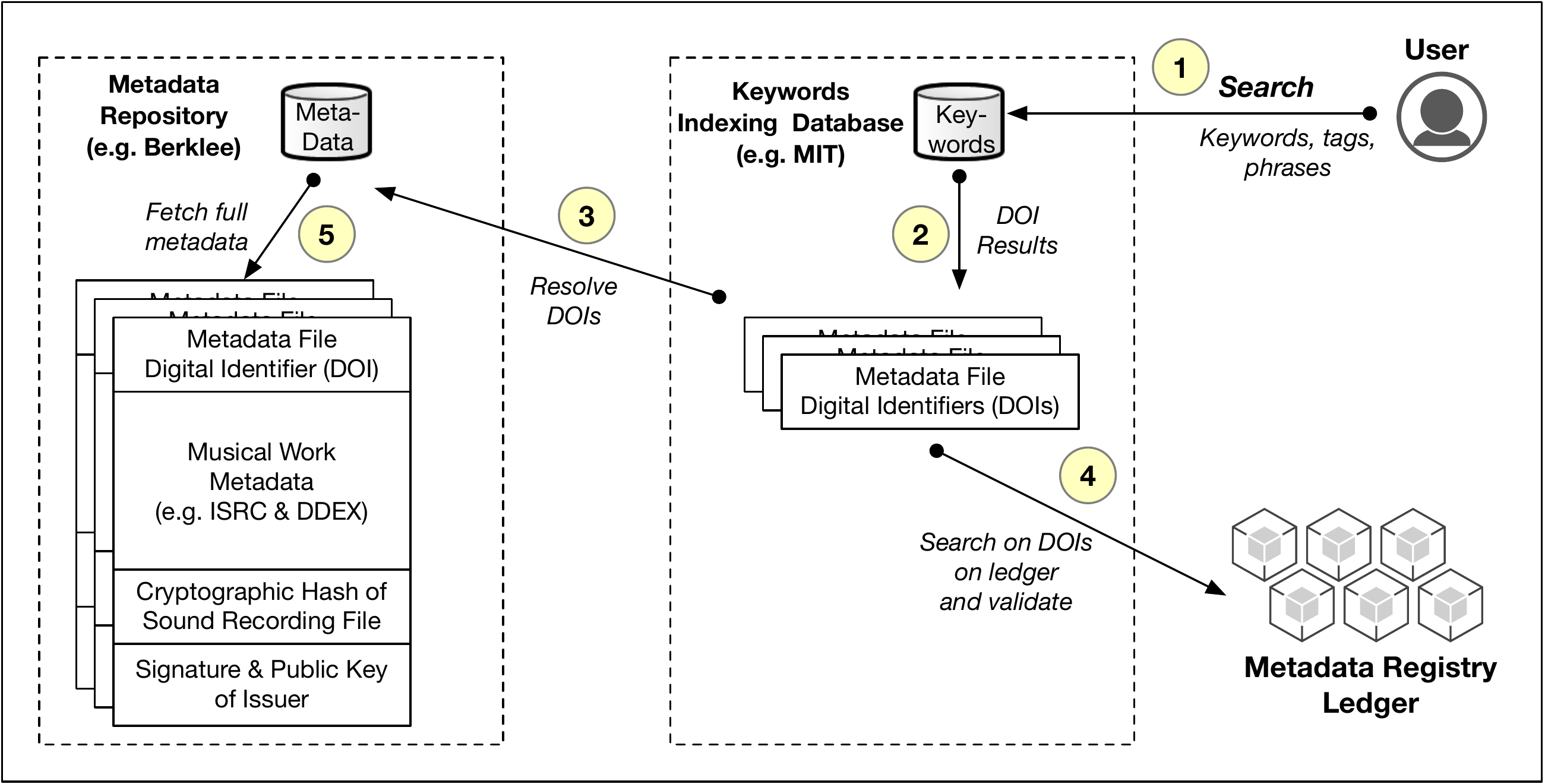}
	%
	% TRIMMING:  trim={<left> <lower> <right> <upper>} and clip options:
	% FULL EXAMPLE: \includegraphics[width=0.4\textwidth, trim={0.5cm 0.5cm 0.5cm 11.3cm}, clip]{image1.pdf}
	%
\caption{Example of search and metadata lookups }
\label{fig:metadata-lookups}
\end{figure}

Figure~\ref{fig:metadata-lookups} illustrates the search process.
In Step~1 a user employs a search-application that performs searches
on local keywords-databases as well as other available globally.
The results of this search in Step~2 is a set of links or DOI values that the
search-application can resolve to the full metadata.
When the search-application presents these results,
the user can chose certain DOIs (e.g. in the user's search-application),
which would result in the search-application fetching 
the full creation-metadata files in Step~3.
In Step~4, the search-application has the option of verifying on the
registry ledger whether 
there exists newer version of the creation-metadata file.

For each creation-metadata file,
the user then can employ the hash of the musical work (e.g. hash of MP3 master file)
found in the creation-metadata 
to fetch a copy of the musical work itself (e.g. MP3 master file) from its location of storage
via a protected API.
This last action will require the user to be authenticated and obtain authorization
from the current owner of the musical work.

Currently, alternative decentralized content management systems
(e.g. Open Index Protocol (OIP)~\cite{OIP-Wiki}) are being developed
to allow the decentralization of the storage of content-files
(e.g. using IPFS~\cite{IPFS})
with a separate locally-cached search terms.
The local caching of search terms (e.g. on the user's computer)
avoids the centralized collection by (and hence user dependence upon)
large search-engine service providers.
This is important because the user's choice of search terms
and keywords may provide valuable insights through social analytics regarding
the user's preference for music genres as well as those
of the user's friends (e.g. people with frequent interactions
in the social interaction-network~\cite{Pentland2015}).

%%%%%%%%%%%%%%%%%%%%%%%%%%%%%%%%%%%%%%%%%%%%%%%%%%%%%%%%%%%%%%%%%%%%%%%%%%%
\section{Future Vision: The 3 Layers of the Global Music Ecosystem}
\label{sec:SummaryConlusions}

In this paper we have discussed the music metadata layer
as the foundation layer for the future global music ecosystem.
This layer is as core to the functioning of the digital music ecosystem
as the DNS infrastructure is core 
to the functioning of the services on the Internet today.
However, in order for new services to grow organically and evolve over time,
we believe that two additional layers (or ``infrastructures'')
will be needed in order for the digital music industry to truly
reach its global market potential.

The three layers or infrastructures of future digital music
are summarized in Figure~\ref{fig:threeinfrastructures}
and consists of the following:
\begin{itemize}

\item {\em Music metadata layer}: This is 
the music metadata layer that we have discussed in the current paper.
This is the bottom-most layer in Figure~\ref{fig:threeinfrastructures}.

There will be several other components of this layer 
that we did not discuss,
such as digital identity management,
cryptographic key management,
protected access to musical works (e.g. sound recording files),
and others.
There is a key role for Artificial Intelligence (AI)
and Machine Learning (ML) technologies at this layer
in solving the music search problem.

\begin{figure}[!t]
\centering
\includegraphics[width=0.9\textwidth, trim={0.0cm 0.0cm 0.0cm 0.0cm}, clip]{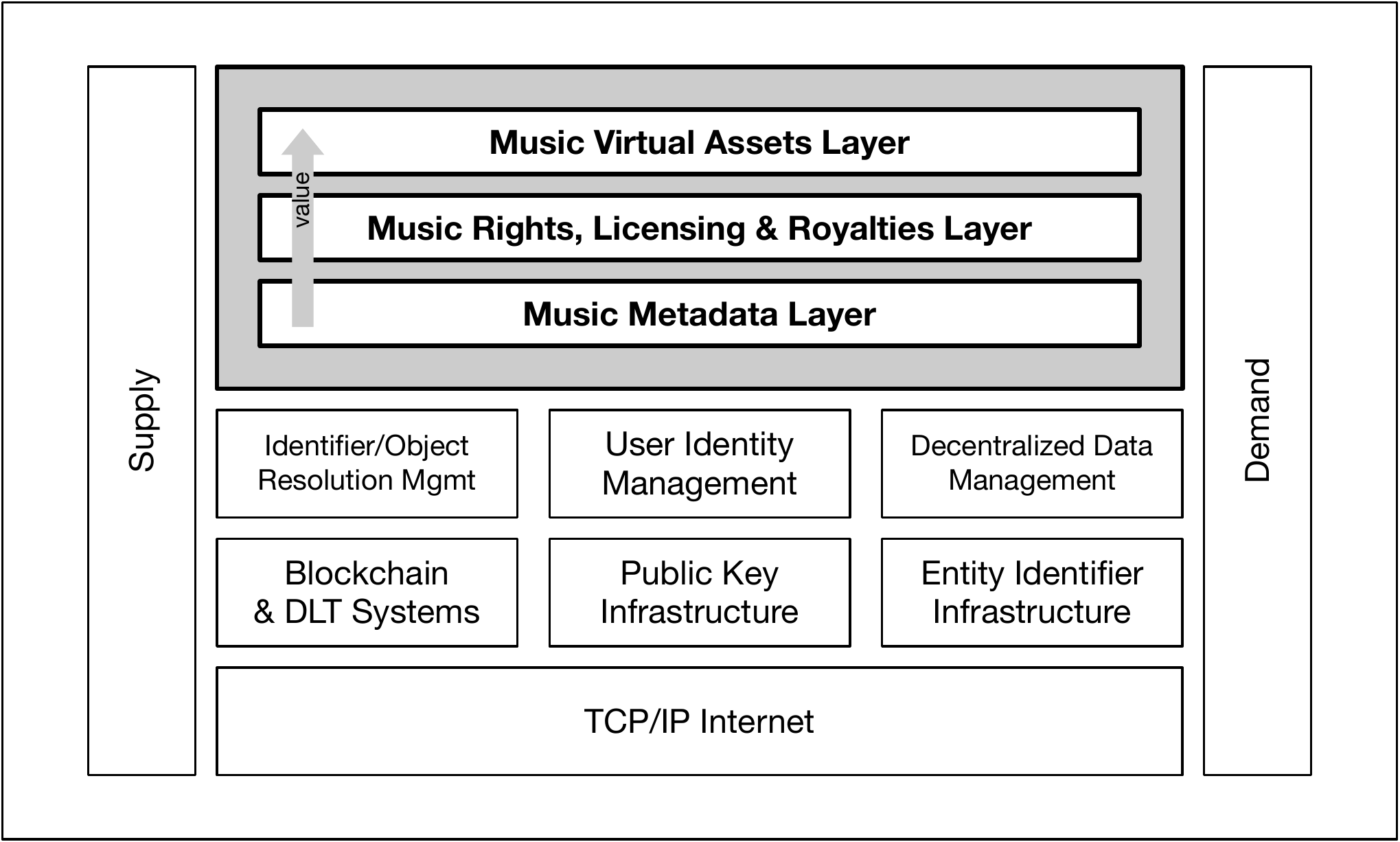}
	%
	% TRIMMING:  trim={<left> <lower> <right> <upper>} and clip options:
	% FULL EXAMPLE: \includegraphics[width=0.4\textwidth, trim={0.5cm 0.5cm 0.5cm 11.3cm}, clip]{image1.pdf}
	%
\caption{The three infrastructures of the future digital music ecosystem}
\label{fig:threeinfrastructures}
\end{figure}

\item	{\em Licensing and Royalties Management layer}:
The second layer needed is one that enables 
a decentralized management of music rights-ownership,
music rights trading (i.e. buy/sell),
license issuance/tracking,
and the collection and distribution of royalties.
This is the middle layer in Figure~\ref{fig:threeinfrastructures}.
Here, we believe that there is a major role for smart contracts technology
to be used to represent business logic
as part of the broader music licensing supply chain management.

It is paramount to recognize that there is a strong dependency of this layer
upon the bottom-most music metadata layer.
Digital licenses cannot be issued by smart contracts (and royalties obtained)
if creation-metadata information is incomplete
or if there are multiple imprecise unauthoritative versions
existing along the music supply chain.
Hence our design principles discussed in Section~\ref{sec:DesignPrinciples}.

One important question for this layer is the confidentially of rights-related
transactions on the distributed ledgers used in this layer.
Research and development continues on cryptographic schemes
(e.g. zero-knowledge schemes) that provide some degree of privacy
to entities that transact on public blockchain networks.
A second key question pertains to the interoperability of 
distributed ledgers and blockchain systems,
something that is severely lacking today~\cite{HardjonoLipton2019a}.

\item	{\em Music virtual assets layer}:
The third layer needed is one that
allows for the recognition of musical works and music-rights
as {\em virtual assets} in the sense of digital tokens~\cite{SEC2019,FATF-Recommendation15-2018}.

A new digital infrastructure is needed to allow 
for the exchange (e.g. buy, sell) of rights in the 
form of digital fungible tokens (e.g. ERC20~\cite{VogelstellerButerin2015}),
or non-fungible tokens (e.g. ERC721~\cite{EntrikenShirley2018}).
Tokens can be used to represent full or partial rights ownerships
of musical works,
and therefore can be used as the basis for distributing royalties
obtained from licensees.

The eventual vision is for this layer to encompass multiple {\em decentralized music-rights trading networks}
which operate on a global scale.
Just like the Internet -- which is composed of multiple ISPs and networks --
the interoperability across trading networks (i.e. distributed ledgers and blockchains)
remains a subject for future research and development in the technology industry.

\end{itemize}
Just as the Internet is not owned by any single entity, organization or country,
we believe that there will be multiple implementations and instantiations of 
the functions and components of the above three layers in Figure~\ref{fig:threeinfrastructures}.
As the history of 1980s Local Area Network (LAN) industry has taught us~\cite{abbate99},
it is futile and even counter-productive
for any single entity to seek 
to own or control entire layers.
As such, technical and operational standards are needed 
to ensure a high degree of interoperability of services 
based on common standardized APIs -- and thus ensure there is significant
competition of services in the market.

%%%%%%%%%%%%%%%%%%%%%%%%%%%%%%%%%%%%%%%%%%%%%%%%%%%%%%%%%%%%%%%%%%%%%%%%%%%
\section{Summary \& Conclusions}
\label{sec:SummaryConlusions}

As mentioned before,
one of the key challenges of the music supply chain today is the lack of consistent,
complete and authoritative information or metadata regarding the creation of a given
musical work.
In this paper we have described the notion of an open access {\em music metadata layer}
that can become the basis for future music-related
transactions on distributed ledgers or blockchain systems.
The metadata layer consists of 
a replicated and decentralized open-access metadata repositories,
which is a series of repositories that stores creation metadata, 
without rights-ownership information and without the copyrighted musical works 
(e.g. composition notes files, sound recording files).
This is coupled with a registry ledger which acts
as a notarization service and which allows anyone in the world to resolve the
identifiers in the registry-metadata 
to one or more copies of the complete creation metadata located on the Internet.
We have described a number of design principles for this music metadata layer.

Our vision is a future music industry that operates globally
based on the three logical layers or infrastructures.
Thus, in addition to the music metadata layer we believe that
a licensing and royalties management layer will be needed
that can automate the business logic processing pertaining to license issuance,
license tracking/accounting, 
rights-ownership trading (i.e. buy/sell), and
royalties collection and distribution.
We believe there is a promising role for smart contracts technologies to
express the various business logic in this layer.
The third and ``uppermost'' layer is one in which 
musical works and music-rights  can be recognized
as virtual assets in the sense of digital tokens.
This tokenization paves the way for 
the evolution towards globally interconnected networks
for the exchange of music virtual assets.

\section*{Acknowledgments}
We thank the various students, staff, artists and creators at Berklee, Lesley and MIT
for their immense interest and support for this project.
We thank especially Heather Reid from the Berklee Library
for her kind help.

~~\\

%\bibliographystyle{IEEEtran}
%\bibliography{IEEEabrv,hardjonobib,thomasrfcbib}


\end{document}